\documentclass[11pt]{asaproc}

\usepackage[ruled,vlined]{algorithm2e}

\usepackage{graphics}
\usepackage{graphicx}
\usepackage{grffile}

\usepackage[utf8]{inputenc}
\usepackage[english]{babel}
\usepackage[backend=bibtex]{biblatex}
\usepackage{csquotes}
\usepackage{tikz}
\usepackage[latex]{xellipsis}
\usepackage{multicol}
\usepackage{amsmath}
\usepackage{floatrow}
\usepackage{mathtools} % for "dcases*"
\usepackage{multirow}
\usepackage{array}
\usepackage{float}
\usepackage{subcaption}
\usepackage{amsthm}
\usepackage{comment}

\newcommand{\red}[1]{\textcolor[rgb]{1,0,0}{#1}} 

\newcommand{\splitatcommas}[1]{\begingroup\lccode`~=`, \lowercase{\endgroup
    \edef~{\mathchar\the\mathcode`, \penalty0 \noexpand\hspace{0pt plus 1em}}%
  }\mathcode`,="8000 #1%
  }

\usetikzlibrary{arrows.meta, chains, positioning,   shapes.geometric}
                
\tikzset{
 disc/.style = {shape=cylinder, draw, shape aspect=0.3,
                shape border rotate=90,
                text width=17mm, align=center, font=\linespread{0.8}\selectfont},
  batch/.style = {shape=rectangle, draw, shape aspect=0.3,
                  align=center, rounded corners, font=\linespread{0.8}\selectfont},
  batch2/.style = {batch, fill=blue!15},
  }

\usepackage{times}

\title{Detection of Data Drift and Outliers Affecting Machine Learning Model Performance Over Time}

\author{Samuel Ackerman\thanks{IBM Research, Haifa; samuel.ackerman@ibm.com} \and  Eitan Farchi\thanks{IBM Research, Haifa} \and  Orna Raz\thanks{IBM Research, Haifa} \and Marcel Zalmanovici\thanks{IBM Research, Haifa} \and Parijat Dube\thanks{IBM Research, Yorktown Heights}}
\begin{document}

\maketitle

\begin{abstract}
A trained ML model is deployed on another `test' dataset where target feature values (labels) are unknown. Drift is distribution change between the training and deployment data, which is concerning if model performance changes. For a cat/dog image classifier, for instance, drift during deployment could be rabbit images (new class) or cat/dog images with changed characteristics (change in distribution). We wish to detect these changes but can't measure accuracy without deployment data labels. We instead detect drift indirectly by nonparametrically testing the distribution of model prediction confidence for changes. This generalizes our method and sidesteps domain-specific feature representation.

We address important statistical issues, particularly Type-1 error control in sequential testing, using Change Point Models (CPMs, \cite{RA12}). We also use nonparametric outlier methods to show the user suspicious observations for model diagnosis, since the before/after change confidence distributions overlap significantly. In experiments to demonstrate robustness, we train on a subset of MNIST digit classes, then insert drift (e.g., unseen digit class) in deployment data in various settings (gradual/sudden changes in the drift proportion). A novel loss function is introduced to compare the performance (detection delay, Type-1 and 2 errors) of a drift detector under different levels of drift class contamination.

\end{abstract}

\section{Introduction
\label{sec:intro}}
The problem of sequential change generally involves detecting when the distribution of data observed over time has changed.  If the problem is phrased as a binary yes/no decision that change has occurred, let us denote cases before the change as `negative' and after change as `positive.'  In this case, deciding change has occurred when it has not (yet) would be a false positive, also known as a Type-1 error.  We address below some of the important statistical issues, such as controlling the false positive rate---that is, the likelihood of falsely detecting drift---at an acceptable level. 
Although change detection is a general problem, we are specifically interested in using it for monitoring changes in the performance (e.g., classification accuracy) of a machine learning (ML) model.  An ML model may serve a critical business or other function, such as classifying loan applicants as good or bad risks, or classifying mammogram images as malignant or benign.  In cases like this, it is important to be able to detect when the model's performance changes (particularly if it degrades) relative to the baseline it was trained on.  In addition, having some realistic statistical guarantee on the decision, since retraining a model may have real costs, is important.  In our experiments, we will couch ML model monitoring in a setting where a new label class is encountered in deployment, which was not present in the training data.  Observations from this new `drift class' should typically cause the model's performance to change.

Some of these practical issues, as well as the main method we will use, change point models (CPMs;  \cite{R15}, \cite{RA12}) have been mentioned in our earlier work in \cite{FDA20} and \cite{AFRZZ19}. Here, we extend this earlier work in two ways.  First, we apply the loss function introduced in \cite{FDA20} to compare the performances of our various change detectors.  Secondly, in addition to detecting changes in the data observed, we would like to present the user with anomalous data examples that have high likelihood of being instances of the new class, to help diagnose the model.

\section{Motivation \label{sec:motivation}}
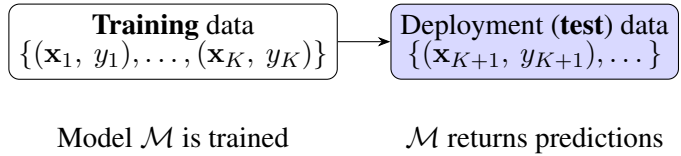
\begin{figure}
\centering
\begin{tikzpicture}[
    node distance = 5mm and 7mm,
      start chain = going right
                   ]
    \begin{scope}[every node/.append style={on chain, join=by -Stealth}]
\node (n1) [batch] {\textbf{Training} data\\              $\{(\mathbf{x}_1,\: y_1),\dots,
            (\mathbf{x}_K,\: y_K)\}$};
\node (n2) [batch2] {Deployment (\textbf{test}) data\\             $\{(\mathbf{x}_{K+1},\:y_{K+1}),\dots\}$};
    \end{scope}
\node[below=of n1]  {Model $\mathcal{M}$ is trained};
\node[below=of n2]  {$\mathcal{M}$ returns predictions};
    \end{tikzpicture}
\caption{\label{fig:training_production}
Illustration of training and deployment data.}
\end{figure}

Our experiments here address the scenario of a ML model $\mathcal{M}$ trained on input data $(\mathbf{x}_i,y_i)$, where $\mathbf{x}_i$ may be of arbitrary domain or dimension, and $y_i$ is a target variable.  Here, we deal with the specific case of classification, but the results are applicable to other prediction tasks.  After training, $\mathcal{M}$ is then deployed on the field on data (Figure \ref{fig:training_production}).  

We would like to monitor $\mathcal{M}$'s performance under deployment and determine statistically if there has been a significant change relative in performance relative to that observed previously.  Such performance changes may be the result of `drift', which  is any change in the \textit{data} ($\mathbf{X}, y$) itself between training and deployment.  The literature on drift often distinguishes between concept and data drift, which deal with changes in the distributions $\textrm{Pr}(\mathbf{Y} \mid \mathbf{X})$ and $\textrm{Pr}(\mathbf{X})$, respectively.  As noted in \cite{MRACH12}, the definitions of these drift types in the literature is inconsistent.  In our experiments, we induce change in the data by altering the distribution of the class labels $\textrm{Pr}(\mathbf{Y})$ observed in deployment; in \cite{MRACH12}, this is termed `prior class probability shift.'  Since we use images from the MNIST image dataset (\cite{LCB99}), and since the image pixel representation should directly depend on the digit class (digits 0 through 9), we change only $\textrm{Pr}(\mathbf{X})$ ---indirectly through changing $\textrm{Pr}(\mathbf{Y})$ ---and not the conditional $\textrm{Pr}(\mathbf{Y}\mid \mathbf{X})$, though this is not an essential aspect of our approach.  Our approach (see Section~\ref{sec:experiments}), which is based on detecting changes in performance metrics of $\mathcal{M}$, should be able to detect any changes in distributions of $\mathbf{X}$ or $\mathbf{Y}$ as long as these are reflected indirectly in changes in the performance metric.

We want our drift detection method to be as generalizable as possible. This includes not making assumptions on the distribution of the input $\mathbf{X}$, including its dimensionality (uni- or multivariate), domain (e.g., image vs numeric data), or on the model type $\mathcal{M}$ used. Hence, rather than modeling the data inputs to detect drift (as in the case of variational auto-encoders, for instance), we will instead monitor a univariate output of the model $\mathcal{M}$ and try to detect changes in $\mathcal{M}$'s performance.  If the data $\mathbf{X}$ or $\mathbf{Y}$ (unobserved) change, but the model performance doesn't, we will probably not detect this.    

We follow a similar approach to that of previous works; for instance, \cite{AFRZZ19}, \cite{KR98}, \cite{L99}, and \cite{LMD11} monitor changes in model confidence, estimated accuracy, or other outputs, since univariate metrics are more easily analyzed by methods such as density estimates and statistical tests. We will show also that many of the methods historically used for change detection suffer from flawed statistical guarantees.

\section{Experimental Setup  \label{sec:experiments}}
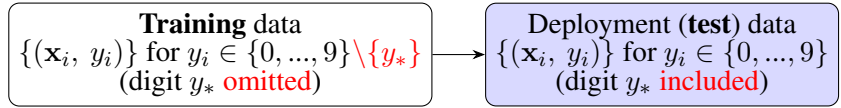
\begin{figure}
\centering
\begin{tikzpicture}[
    node distance = 5mm and 7mm,
      start chain = going right
                   ]
    \begin{scope}[every node/.append style={on chain, join=by -Stealth}]
\node (n1) [batch] {\textbf{Training} data\\              $\{(\mathbf{x}_i,\: y_i)\}$ for $y_i\in \{0,\xelip,9\}\red{\setminus\{y_*\}}$\\
 (digit $y_*$ \red{omitted})};
\node (n2) [batch2] {Deployment (\textbf{test}) data\\             $\{(\mathbf{x}_i,\: y_i)\}$ for $y_i\in \{0,\xelip,9\}$\\
 (digit $y_*$ \red{included})};
    \end{scope}
%\node[below=of n1]  {Logistic regression $\mathcal{M}$ is trained};
    \end{tikzpicture}
\caption{\label{fig:mnist_omitted_class}
Scenario for training and prediction for MNIST data with omitted digit $y_*$ (drift class) in the deployment data.}
\end{figure}

The experimental data used here derives from experiments from \cite{AFRZZ19} on MNIST (\cite{LCB99}) image data, which consists of handwritten images of digits 0 through 9 (ten classes); let these classes be denoted $\mathbf{y}=\{0,1,\dots,9\}$.  In \cite{AFRZZ19}, for each digit class $y_*\in \mathbf{y}$, a training set is formed which contains images from the remaining digits $\overline{y_*}=\mathbf{y}\setminus y_*$.  A logistic regression model $\mathcal{M}$ is trained on this subset (with $y_*$ omitted), and hence can only classify images as belonging to one of these nine classes.  Then, $\mathcal{M}$ is presented with images of the same classes of the training classes $\overline{y_*}$, along with images from the omitted digit class $y_*$ (see Figure~\ref{fig:mnist_omitted_class}).  Images where $y_i=y_*$ are called `drift class' since it is a class not observed in training.  In \cite{AFRZZ19}, experiments were also done where the drift class observations differed from the training data by being either a random noise or an out-of-domain (non-digit) image, but here we use only the results where the drift class was an unseen digit.

As mentioned in Section~\ref{sec:motivation}, we will monitor a model output; here, this is the outputted model confidence, denoted $z\in [0,1]$, of the most likely predicted class $\hat{y}$ for a given instance.  For the experiments that follow,  we use only fixed sets $\{(z_i, y_i)\}$ outputted from the MNIST experiments of \cite{AFRZZ19}, where $z_i$ is the outputted confidence on a given prediction $\mathcal{M}(\mathbf{x}_i)$ for which $y_i$ is the true class.  In order to demonstrate the performance of our drift detector, these data are held fixed and we repeatedly sample from them in the following procedure.

\begin{figure}
    \centering
    \begin{tikzpicture}[
    node distance = 5mm and 5mm,
      start chain = going right
                   ]
    \begin{scope}[every node/.append style={on chain, join=by -Stealth}]
\node (n1) [batch] {Batch 1\\                             $\{z_1,\xelip,z_{20}\}$};
\node (n2) [batch] {Batch 50=$B$\\                  $\{z_{981},\xelip,z_{1,000}\}$};
\node (n3) [batch2] {Batch 51=$B+1$\\                 $\{z_{1,001},\xelip,z_{1,020}\}$};
\node (n4) [batch2] {Batch 100\\                         $\{z_{1,981},\xelip,z_{2,000}\}$};
\end{scope}
\end{tikzpicture}
    \caption{\label{fig:batch_diagram} Construction of batches of prediction confidences $z_t$.}
\end{figure}
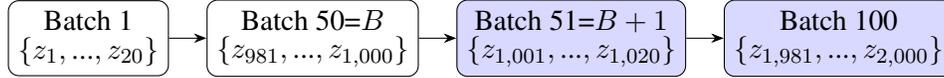

For a given digit $y_*\in \mathbf{y}$ (say, `0'), do the following:
\begin{itemize}
    \item The relevant experiment from \cite{AFRZZ19} is that where the remaining digit classes $\overline{y_*}$ were trained on.
    \item Form a sequence of 100 batches of 20 values each of outputted confidences $z_t$ (2,000 values total).  Batch $j\in \{1,\dots,100\}$ thus consists of $\{z_{20j-19}, z_{20j}\}$.  See Figure~\ref{fig:batch_diagram}.  Let $b(t)=\lceil t/20\rceil$ be the batch index of an observation index $t$.
    \item Batches 1--50 consist of values of $z_t$ where the true digit label $y_t$ is one of the training digit classes (e.g., 1--9) (but not the same training instances themselves).
    \item Batch $j$ has $100p_j\%$ of values consisting of $z_t$ where $y_t=y_*$, the omitted class.  For $j=1,\dots,50$, $p_j=0$, by definition; for $j=51,\dots,100$, $0\leq p_j \leq 1$. 
\end{itemize}

The changepoint $K=1,000$ (the last index in batch 50) is the last sequence index $t$ before which the drift class ($z_t$ for $y_t=y_*$) may be inserted.  For conciseness, let $B=b(K)=50$, the batch of the changepoint.  The next observed $z_{1,001}$ has probability $p_{51}$ (since $b(1,001)=51$) of being from the omitted class $y_*$. The goal is to detect, at some time index $d>K$, that a change happened at some $t<d$.  A false alarm occurs if $d\leq K$, that is a (false) `detection' is made before it actually happened.  We wish to avoid false alarms (controlling its likelihood below a pre-determined level $\alpha$, such as 0.05) while simultaneously detecting as quickly as possible, that is minimizing the delay $d-K>0$.  The above procedure is repeated 50 times for each digit $y_*$ and the results aggregated.  For instance, the estimated Type-1 error (false alarm probability) will be the number of repetitions for which $d\leq K$, divided by 50.

\section{Sequential Change Detection  \label{sec:sequential}}
Many previous methods to detect concept drift have tended to demonstrate their success by carrying out \textit{repeated} statistical hypothesis tests (or threshold checks) on some data value or statistic (function) of it, observed over time.  In general, a statistical hypothesis test formulates a null distribution---possibly nonparametric, described by an empirical set of values---against which a data statistic is compared and a measure of statistical significance (such as a p-value) is calculated.  For instance, \cite{LMD11} calculate a signal value on the data, and count how often the signal value is observed to be a given number of standard deviations above the mean, calculated historically.  They consider several rules for how many times this has to happen ($x$) in the past ($y$) observations (`Western Electric Rules') in order to declare drift.  In \cite{SK17}, an intriguing method is presented to detect ML model potential failure by monitoring a model's ``regions of uncertainty" through its ``marginal density".  The thresholds for change detection are based on the  empirical mean and standard deviations of their metric under $k$-fold cross validation.  While the experimental results are adequate, they do not appear to attempt to control the false positive rates either sequentially or using the criteria of \cite{RATH12}.

It is well-known that repeated statistical testing without appropriate adjustment will cause unacceptable rates of false alarm (Type-1 errors). In the drift setting, the null hypothesis ($H_0$) will be that drift has not occurred, while the alternative ($H_A$) is that it has.  Consider a test for identifying drift, which has decision threshold $\alpha$ (e.g., 0.05).  If applied on a single batch of non-drifted data ($H_0$), this test will raise a false alarm (falsely say there is drift) with probability $\alpha$.  This $\alpha$ is a parameter specified by the user.  However, say the test is applied to $w$ independent windows of non-drifted data, and that the drift will be detected at the \textit{first} batch that seems to have drifted (that is, its p-value is $\leq\alpha$).  This is equivalent to a drift detection if  \textit{any} one of the $w$ batches appears to have drifted significantly.  The probability of the correct decision here (that is, of none of the windows falsely indicating drift) is now $(1-\alpha)^w$, rather than the higher $1-\alpha$ for a single test.  Thus, a test naively applied multiple times without proper adjustment will not give the expected $\alpha$-level statistical guarantee.

The rationale for making repeated drift tests in our setting is that we want to detect drift as soon as possible, and hence have to conduct the test at multiple time points without waiting to observe all the data; in many cases, the data may be an `infinite' stream without a predetermined sample size.
The problem illustrated above of false alarms under multiple tests is compounded when the data windows are not independent, as in the case when we want to examine overlapping windows of historical values.

For instance, consider an independent normal sample $\{x_1,x_2,\dots,x_{100}\}\sim \textrm{ iid } N(\mu=0,1)$.  For the 81 values $t=20,21,\dots,100$, let $\mathbf{x}_t=\{x_1,\dots,x_t\}$.  Let the pair of hypotheses be $H_0\colon \: \mu=0$ and  $H_A\colon \: \mu\ne 0$; the null hypothesis $H_0$ is always true since the true mean is unchanged at $\mu=0$.  The hypothesis test is repeatedly performed for each of the overlapping sets $\mathbf{x}_t,\:t \geq 20$.  The first such sample whose p-value $<\alpha$ for a pre-chosen $\alpha$---that is, the first time the sample mean $\bar{\textbf{x}}_t$ appears significantly different from $\mu=0$--- triggers a decision of change.  Let $V$ denote the number of these change detections out of the 81 tests done. For the standard case of $\alpha=0.05$, the probability of at least one (false) detection $V$, that is, the expected false alarm probability of this entire procedure, is around 23\% rather than 5\% ($\alpha$); see Table~\ref{tab:peeking_problem}. In the independent (non-overlapping set) case, the false alarm rate would be $1-(1-\alpha)^{81}$, which is $\approx98.4\%$ for $\alpha=0.05$, almost a certainty. Thus, the false alarm rate of the overlapping case is not as bad as the independent case, but it is still much higher than the desired $100\alpha\%$.

\begin{table}[ht]
\centering
\begin{tabular}{r|rrrr}
  \hline
$\alpha$ & 0.05 & 0.01 & 0.005 & 0.001 \\ \hline
  $\textrm{Pr}(V\geq 1)$ & \bf{0.2296} & 0.0678 & 0.0360 & 0.0074 \\ 
  $\textrm{E}(V)$ & 4.0825 & 0.7680 & 0.3598 & 0.0527 \\ 
  \hline
\end{tabular}
\caption{\label{tab:peeking_problem} Observed false alarm probability for repeated testing on overlapping samples $20,\dots,100$, based on 10,000 simulations.}
\end{table}

As noted by \cite{RATH12} in presenting their ECDD method for detecting concept drift in Bernoulli-distributed variables (e.g., binary indicator of correct classification), many drift detection methods suffer a weakness in that they cannot properly demonstrate that the false alarm probability is controlled in reality under a wide variety of settings.  The average run length (denoted $\text{ARL}_0$) is the average number of observations between false positives.  For instance, to use their example, say the creators of a given method experimentally demonstrate that their method makes one false alarm detection every 100 observations ($\text{ARL}_0=100$).  Because the rate of positive instances (instances of drift) in the data stream is unknown, such a rate of mistaken decisions may seem low in a different application but will likely be too high for a practical sequential detection setting if a detection of drift has potentially costly consequences, such as forcing model retraining or examination of the data, and we are to trust its decision.

A stricter criterion, such as one false positive every 5,000 observations rather than 100 (higher $\text{ARL}_0$) may be needed to demonstrate that a positive decision of drift is to be trusted.  It seems that many methods of drift/shift detection may thus use data streams that are too short relative to what is likely to be encountered in a realistic scenario; this is particularly true if a method must make a single positive decision and be evaluated based on it (i.e., `all-or-nothing' scoring) rather than being allowed to make more than one false positive and being evaluated on the average rate.

The sequential statistical technique we adopt to deal with these issues is the Change Point Model (CPM) developed by \cite{RA12} and implemented as the \texttt{cpm} package \cite{R15} for \texttt{R} software. The CPM allows us to conduct repeated backwards-looking drift detection while controlling false alarm probability (Type-1 error) for a user-desired value of $\textrm{ARL}_0$; this method also has theoretic statistical guarantees on correctness, not just limited experimental results, as we summarize below.  We note also that two of the authors of the CPM (Adams and Ross) were authors of the ECDD \cite{RATH12} discussed above.

\begin{figure}[ht]
    \centering
    \includegraphics[scale=0.6]{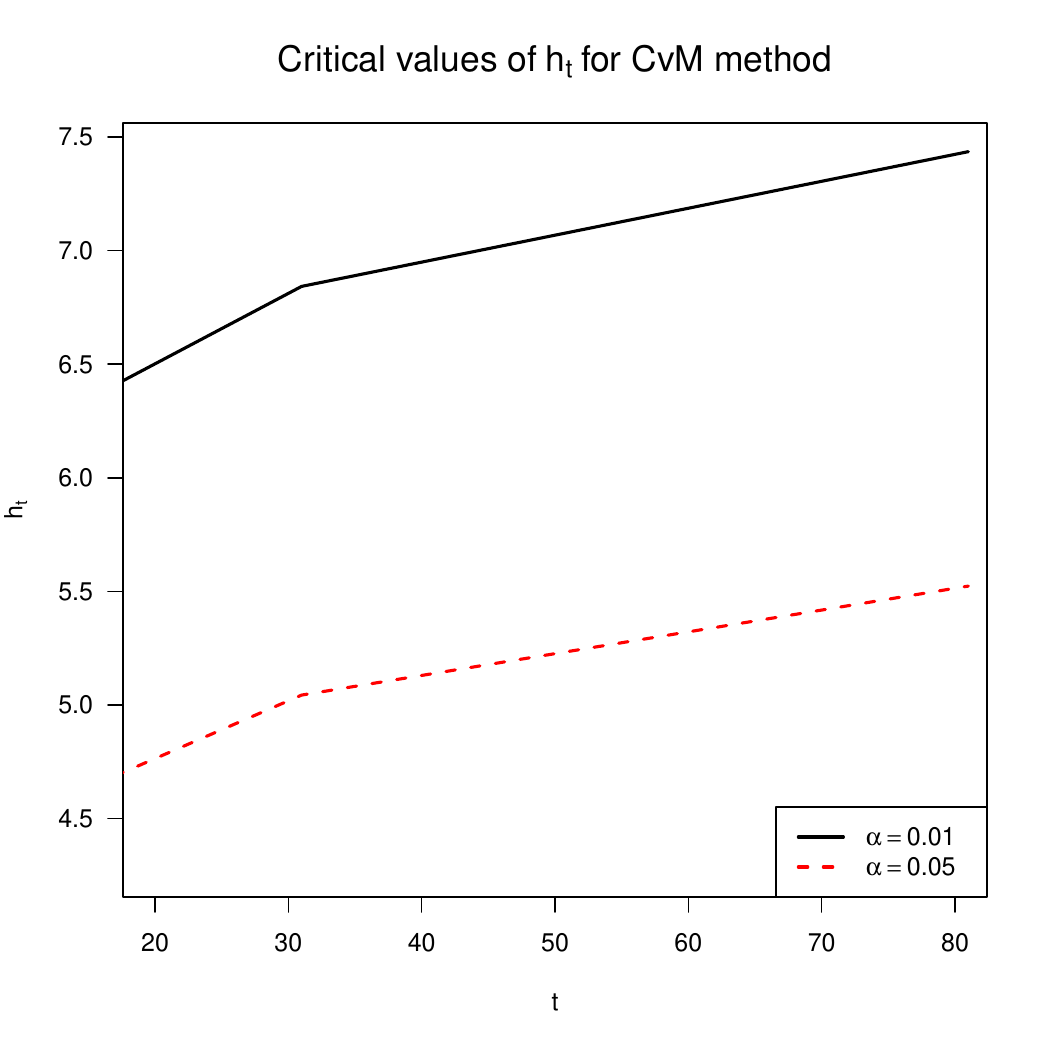}
    \caption{\label{fig:h_t} Critical values $h_t,\: t=20,\dots,80$ for the Cramer-von-Mises CPM for $\alpha=0.01, \:0.05$.  For a lower $\alpha$ (higher significance), the critical values are higher.  Also, for each $\alpha$, $h_t$ increase with $t$.}
\end{figure}

Algorithm~\ref{alg:cpm} shows the outline of the CPM method; the interested reader is directed to \cite{RA12} for full details.  Here, data $x_1,x_2,\dots$ (in our application, the inputs are the observed confidence values $z_t$) are input.  At each time point $t$, the data are split at each potential time $k=2,\dots,t-1$ into before/after samples $\{x_1,\dots,x_k\}$ and $\{x_{k+1},\dots,x_t\}$.  These samples are then input into a two-sample goodness-of fit test such as Cramer von-Mises, Student T-test, or Kolomogorov-Smirnov; the availability of many tests and the nonparametric options make this method attractive.  The goodness-of-fit test yields a set of statistics $\{W_{k,t}\}_{k=2}^{t-1}$, one for each split, calculated using a given function $\textrm{Diff}(\cdot)$, which are then normalized.  At each time $t$, let $\tau=\textrm{argmax}_k\: W_{k,t}$ be the index $k$ with the largest corresponding statistic $W_{\tau,t}$; $\tau$ represents the guess of the true changepoint, denoted $K$ (i.e., $\tau=\hat{K}$ based on the data observed until then). $W_t=W_{\tau,t}$ is then compared to a decision threshold $h_t$, and if it exceeds $h_t$, change is declared at time $d=t$.  This is based on the fact that the outputted statistic of a goodness-of-fit test is maximized if the proposed split point is actually the correct one, meaning the two samples compared will be the most different they can be.  The key point is that these thresholds $h_t$ increase over time (see Figure~\ref{fig:h_t}) rather than staying constant, so that the the probability of of a false alarm is always the desired $\alpha$, given that one has not yet declared change has occurred; that is,

\begin{itemize}
    \item $\textrm{Pr}(W_1>h_1\mid \textrm{ no change})=\alpha$
    \item $\textrm{Pr}(W_t>h_t\mid \textrm{ no change by $t$ and } W_{t-1}\leq h_{t-1},\dots,W_1\leq h_1)=\alpha, \:\forall t>1$
\end{itemize}

The estimated changepoint is thus $\hat{K}=\tau$ when $t=d$, and its batch is $\hat{B}=b(\hat{K})$.  A true detection is when $\hat{K}>K$, which requires in addition that $b(d)>B=b(K)$, since true detection can happen only at the first instance of drift or later.

\SetArgSty{textnormal}
\begin{algorithm}[ht]
\SetAlgoLined
\KwResult{DRIFT detected}
 //Set your significance level \\
  $\alpha  = 0.05$;\\
 //Initial stabilization period\\
  $t_0=25$;\\
  $t=t_0$; \\
  $d=\infty;\: \hat{K}=\infty$;\\
 //Sequential critical values\\
 $h_b,\:h_{b+1},\dots$\\
 DRIFT = \textbf{False}

 \While{\text{DRIFT} = \textbf{False}}{
    //Consider each possible split of data into two before/after subsets\\
    \For{$k=2,\dots,t-1$}{
        $\textbf{X}_0=\{x_1,\dots,x_k\}$\\
        $\textbf{X}_1=\{x_{k+1},\dots,x_t\}$\\
        $W_{k,t} = \text{norm}(\text{Diff}(\textbf{X}_0,\textbf{X}_1))$\\
    }
    //Find most significant split\\
    $\tau=\text{argmax}_k\: W_{k,t}$\\
    %$D_t=D_{\hat{\tau},t}$\\
    \If{$W_{\tau,t} > h_t$}{
       $d=t;\: \hat{K}=\tau$;\\
       DRIFT = \textbf{True}
    }
    $t$++;
    
    }
    \Return $d$ (detection time), $\hat{K}$ (estimated changepoint); both are $\infty$ if no detection
 \caption{\label{alg:cpm}Outline of Ross \& Adams' CPM algorithms}
\end{algorithm}

\section{Drift scenarios  \label{sec:scenarios}}
\begin{figure}
    \centering
    \includegraphics[scale=0.4]{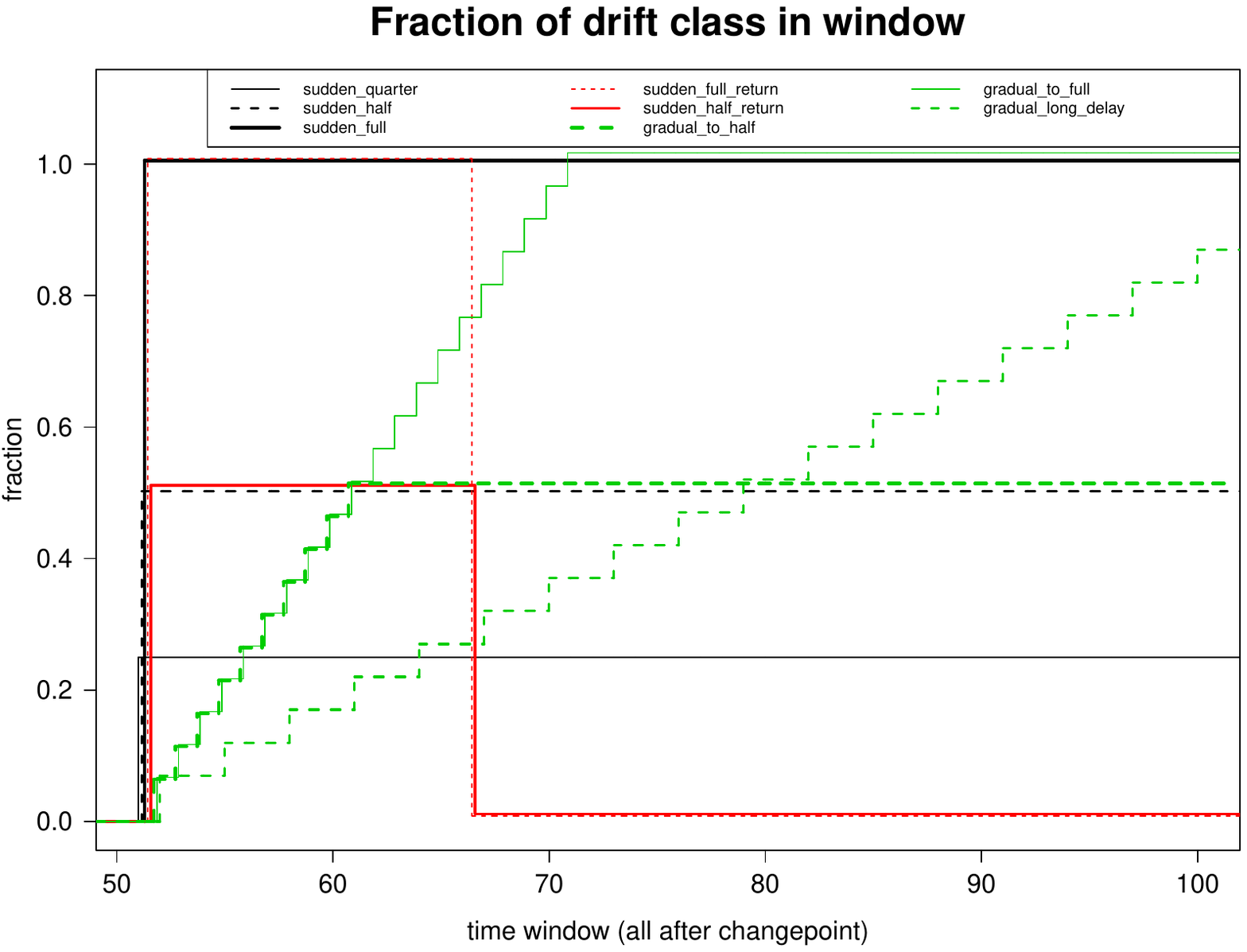}
    \caption{\label{fig:drift_scenarios} Drift scenarios in terms of drift (omitted digit) class proportion $p_j$ in batches $j=51,\dots,100$.  See Section~\ref{scenario_desciptions}.
}
\end{figure}

Many change detection methods have distinguished between change that happens gradually vs abruptly; \cite{KR98} for instance, use the term `drift' for gradual and `shift' for abrupt change.  In our application, gradual change can be thought of as gradual (slow increasing) introduction of the omitted digit class in the data stream.  In Section~\ref{sec:experiments}, we noted that $p_j$ is the omitted digit class proportion in batch $j$.  We first propose to test our application of the CPM under various drift scenarios (see Figure~\ref{fig:drift_scenarios}) as follows;
after each scenario name, in parentheses, appears an abbreviation that appears in the plots Figure~\ref{fig:false_alarm} and after.
Note that $p_j=0,\:\forall j\leq50$ under all scenarios:

\label{scenario_desciptions}
\begin{itemize}
    \item `Sudden' changes where the contamination rate in the first post-change batch $j=51$ remains constant for the entire history, that is $p_{51}=\dots=p_{100}$.
    \begin{itemize}
        \item \textbf{sudden\_quarter} (S\_25\%): $p_j=0.25,\: j\geq 51$.
        \item \textbf{sudden\_half} (S\_50\%): $p_j=0.50,\: j\geq 51$.
        \item \textbf{sudden\_full} (S\_100\%): $p_j=1.0,\: j\geq 51$.
    \end{itemize}
    \item `Sudden' changes where the drift classes disappear after a certain time period.
        \begin{itemize}
        \item \textbf{sudden\_half\_return} (SR\_50\%): $p_j=0.5,\:  51\leq j \leq 65$; $p_j=0,\: j \geq 66$.
        \item \textbf{sudden\_full\_return} (SR\_100\%): $p_j=1.0,\: 51\leq j \leq 65$; $p_j=0,\: j \geq 66$.
    \end{itemize}
    \item Gradual changes
    \begin{itemize}
        \item \textbf{gradual\_to\_half} (G\_50\%): $p_j=0.05(j-50),\: 51\leq j \leq60$; $p_j=0.50,\: j \geq 66$.
        \item \textbf{gradual\_to\_full} (G\_100\%):  $p_j=0.05(j-50),\: 51\leq j \leq 70$; $p_j=1.0,\: j \geq 71$.
        \item \textbf{gradual\_long\_delay} (G\_LD): $p_j=0.05\lceil(j-50)/3\rceil,\: j \geq 51$ ($p_j$ increases by 5 ppt every 3 batches).
    \end{itemize}
\end{itemize}

The idea is that sudden changes in $p_j$ should be easier to detect than gradual changes that begin smaller and reach the same level, and changes that persist should be easier to detect than changes that return to the pre-change situation ($p_j=0$).  The CPM has options to include as its goodness-of-fit test the Student-T and Cramer von-Mises (CvM) test, among others.  Using a fixed performance metric---principally detection delay, and the  false and missed alarm probabilities---we can compare a given CPM detector to others, and compare the results of a given detector for the different drift scenarios.  However, how can we compare two detectors' success in several scenarios over these different metrics?  Is it better to have a false alarm probability of 0.05 and delay of 7 batches, or a probability of 0.03 and delay of 10 batches?

In \cite{FDA20} we presented a novel loss function.  Recall, $K$ is the true changepoint (1,000 in our setting); $b(t) = \lceil t/20\rceil$ is the batch of a time index $t$, and $B=b(K)=50$, the last batch before the change; and $\{p_j\},\: j=1,\dots,100$ are the contamination values associated with a given drift scenario, where $j$ is the batch index. Let $d\in \{1,\dots,100\}$ be the window of detection, if, made, where $b(d)=51$ is the best result; if no detection is made, let $d=\infty$. Let $L_0, L_1<0$ be constants specifying a penalty incurred for a false alarm (Type-1 error or false positive) and missed alarm (Type-2 or false negative), respectively.  A given result ($K, d,\{p_j\}$), which together describe the true changepoint, detected time, and drift scenario, receives a loss score defined as

\[
L(K, d,\{p_j\})=
\begin{dcases}
L_0& \text{if } d\leq K\\
L_1& \text{if missed ($d=\infty$)} \\
L_1 -\left(\frac{ L_1}{\prod_{j=b(K)+1}^{b(d)}\: (1+p_j)^{\frac{b(d)-j}{b(d)-b(K)}}}\right) & \text{if } K<d<\infty \\
\end{dcases}
\]

Setting $L_0<L_1$ (we set $L_0=-1000$ and $L_1=-250$) means a false alarm is penalized more than a missed detection, forcing the detector to be more conservative.  The best result is if the detection at $d$ falls in batch $b(d)=b(K)+1=51$, the first batch after the change.  If this happens, the denominator of the third expression in $L(\cdot)$ is 1 since the exponent is 0.  For true detections where $b(K)<b(d)<\infty$, the loss function penalizes delayed detection by compounding $p_j$ for each batch $j$ for the number of windows it has been present until the detection.  Since $L_0$ factors out of the third expression, all that matters is the ratio of $L_0$ and $L_1$, not their absolute values.

\begin{figure}
    \centering
    \includegraphics[scale=0.4]{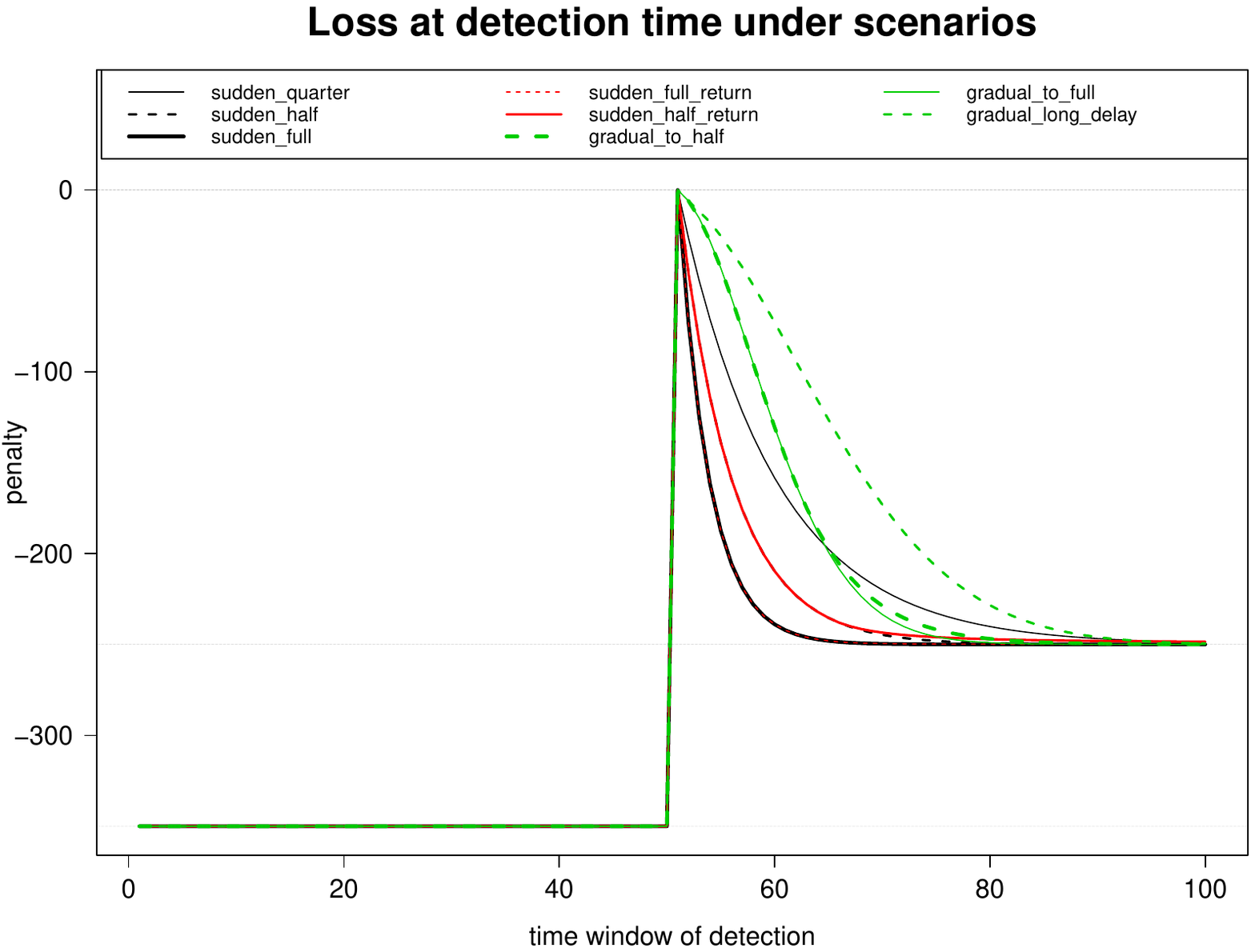}
    \caption{\label{fig:drift_scenarios_loss} Loss function value for $L_0=-350$ and $L_1=-250$.
}
\end{figure}

This is illustrated in Figure~\ref{fig:drift_scenarios_loss}; we set $L_0=-350$ here for clarity.  For detection batch $b(d)\leq 50$, the loss is constant at $L_0$.  For $b(d)=51$, the loss is 0 since this is optimal.  For $b(d)>51$, the loss declines monotonically to $L_1=-250$.  However, for instance, $\textbf{sudden\_full}$ (thick solid black line), the loss grows more negative faster than, say, for $\textbf{sudden\_quarter}$ (thin solid black line), where the drift contamination is smaller.  Thus, for instance, at any given detection time $b(d)$, the first receives lower loss; also, the same given loss value (y-axis) is incurred earlier by the first than the second.  The most gradual and lowest contamination $\textbf{gradual\_long\_delay}$ (thin dashed green line) scenario has loss that grows the slowest.

\section{Outlier Detection \label{sec:outlier_detection}}
In addition to detecting drift in terms of a change in distribution of the confidences $z_t$, we may also want to provide the user suspicious instances of the underlying data $(\mathbf{x}_t, y_t)$, even though only $\{z_t\}$ are used in the detection itself.  The hope is that these suspicious instances will help the user understand the cause of the drift and make appropriate adjustments to $\mathcal{M}$, or possibly, the detection procedure itself.

As mentioned, the CPM (Algorithm~\ref{alg:cpm}) makes a detection at time $t=d$ and returns an estimate $\hat{K}$ of the true changepoint $K$ (=1,000).  Thus, the pre-drift and post-drift samples are $\mathbf{Z}_0=\{z_1,\dots,z_{\hat{K}}\}$ and $\mathbf{Z}_1=\{z_{\hat{K}+1},\dots,z_d\}$.  Since our detection is in change of distribution of $z_t$, a natural solution is to select the most anomalous confidence values in $\mathbf{Z}_1$ relative to $\mathbf{Z}_0$.  Define $\mathbf{\Theta}=\{t\colon\: t\in \{\hat{K}+1,\dots,d\}\;\text{s.t.}\; z_t\;\text{is an outlier relative to}\;\mathbf{Z}_0\}$, so the anomalous confidence values are $\{z_t\}_{t\in \mathbf{\Theta}}$.  The user will thus be presented with $\{(\mathbf{x}_t,\: y_t)\}_{t\in \mathbf{\Theta}}$ in the original domain of the data (e.g., images and class label) as potential outliers.  We may additionally want to restrict $|\mathbf{\Theta}|$ to be of maximal size, say 7; this is based on the ideas introduced in \cite{M56} that human short-term memory can handle only a limited number of items.  We do not think all of $\{(\mathbf{x}_t,\: y_t)\}_{t\in \mathbf{\Theta}}$ will be outliers, but rather that they include at least one (and at higher proportion than a random sample from $\{(\mathbf{x}_t,\: y_t)\}_{\hat{K}+1\leq t\leq d}$).  Nor do we hope to return most of the drift instances to user, but rather a few representative ones that will help in diagnosis.  It is assumed that the user should be able to diagnose the problem relatively easily once pointed to these items, such as seeing by visual inspection that the digit is a new class, or that the handwriting style or image resolution has changed.

We use the two-sample local kernel density-based test, proposed in \cite{D13}.  It is implemented in \texttt{R} package \texttt{ks} (kernel smoothing; \cite{DWCG18}) as \texttt{kde.local.test} for the general case and \texttt{kde.local.test.1d} for  the simpler univariate case.  In the univariate case, the test takes as input two univariate samples (here, $\mathbf{Z}_0$ and $\mathbf{Z}_1$) and conducts a nonparametric kernel density estimate on each, denoted $\hat{f_0}$ and $\hat{f_1}$, a process which includes determining the optimal bandwidth values for each, which are inputs into the algorithm.  The domain (in our case $z_t\in [0,1]$) is discretized into $n$ equally-spaced points $\{\delta_1,\dots,\delta_n\}$ where $\delta_1=0$ and $\delta_n=1$.  Let $\Delta_i=\hat{f_1}(\delta_i)-\hat{f_0}(\delta_i)$.  Assuming this discretization represents the densities well enough, a local test will be done at each $\delta_i$ to see if the difference $\Delta_i$ between the two densities is significantly different from 0.  At each $\delta_i$, the statistic $\chi^2_{\text{stat}}=\left(\displaystyle{\frac{\Delta_i}{\widehat{SD}(\Delta_i)}}\right)^2$ is calculated; the standard deviations of the differences $\Delta_i$ are obtained from a formula that relies on the density bandwidths.  Each statistic is independently chi-squared $\chi_1^2$ distributed, since they are squared normally-distributed random variables.  For each, the upper-tailed p-value is calculated;
given a desired threshold $\alpha_*$ (which, in our application, may be different from the $\alpha$ of the CPM), a Hochberg multiple-testing adjustment is applied to them, yielding adjusted p-values $\{\pi_1,\dots,\pi_n\}$, and thus a decision as to which $\delta_i$ are locations of significant differences.

Now, the individual tests can be used to determine significant regions.  Let $\delta_a<\delta_b$ be two points in the discretization.  We can say \textit{an interval} $[\delta_a,\delta_b]\in [0,1]$ is an area where $f_0$ and $f_1$ differ significantly if $\pi_i<\alpha_*,\:\forall a\leq i\leq b$, that is if there is a significant density difference at all intermediate discretization points $\delta_i$.  Let $\Lambda$ be the union of all such intervals, if they exist. Thus, we may specifically define $\mathbf{\Theta}=\{t\colon\: t\in \{\hat{K}+1,\dots,d\}\;\text{s.t.}\; z_t\in\Lambda\}$, again restricted to a maximal size and choosing the intervals in $\Lambda$ in order of the significance of their p-value given it is below $\alpha*$.
The user is thus presented with $\{(\mathbf{x}_t,\: y_t)\}_{t\in \mathbf{\Theta}}$ as before.

For the purposes of this application of outlier detection, we only consider those points $\delta_i$ for which $\pi_i <\alpha_*$ and specifically $\hat{f_1}(\delta_i) > \hat{f_0}(\delta_i)$, rather than simply having them be significantly different.  That is, we consider only one tail ($>$) of the two-tailed ($\ne$) results.  Considering only intervals $[\delta_i,\delta_j]$ where post-change confidences $z_t$ are \textit{more concentrated} thatn pre-change ($\hat{f_1} > \hat{f_0}$) rather than less-concentrated, is more likely to identify unusual observations. It turns out that the model $\mathcal{M}$'s predictions on the drift class observations can actually tend to be \textit{more confident} (incorrectly) than less confident.  This means that the intervals $[\delta_i,\delta_j]\in [0,1]$ that we identify as potentially containing outliers can occur at the upper (confident) edges of [0,1] in addition to the lower ones.

\begin{figure}
    \centering
    \includegraphics[scale=0.4]{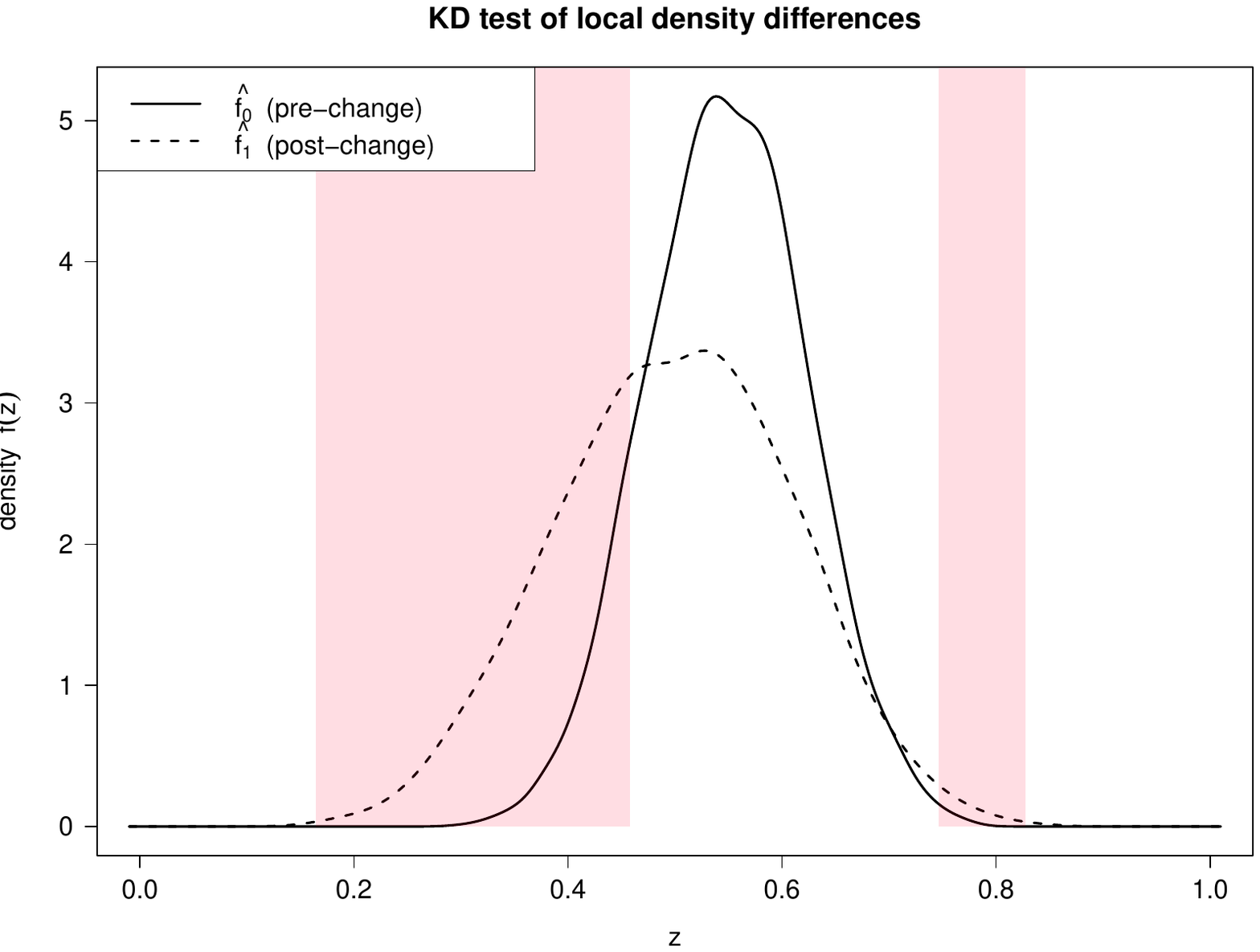}
    \caption{\label{fig:local_density_differences} Illustration of kernel density test of local differences between two Beta-distributed samples $f_0\sim \mathcal{B}(\alpha=24, \:\beta=20)$ and $f_1\sim \mathcal{B}(\alpha=10, \:\beta=10)$.  The red shaded areas show areas of the domain $z\in[0,1]$ where $\hat{f_1}(z)>\hat{f_0}(z)$ significantly.  The regions are formed by joining together all individual points $\delta_i\in[0,1]$ in the discretization of the domain where $\hat{f_1}(\delta_i)>\hat{f_0}(\delta_i)$.
}
\end{figure}

Figure~\ref{fig:local_density_differences} shows an example of this local test on two Beta-distributed samples with densities $f_0\sim \mathcal{B}(\alpha=24, \:\beta=20)$ and $f_1\sim \mathcal{B}(\alpha=10, \:\beta=10)$. The red shaded area represent the areas where $\hat{f_1}>\hat{f_0}$ significantly.  Note that, for instance, $\hat{f_1}>\hat{f_0}$ to the right of the leftmost shaded region, but that this height difference is not judged as statistically significant.  The outlier sample is indices $t\in\mathbf{\Theta}$ where $z_t $ fall in the shaded regions, from which we only select the most anomalous (regions with lowest p-values) subject to the limited number of outliers (10) we wish to present. 

\section{Results
\label{sec:results}}
As mentioned in Section~\ref{sec:experiments}, for each digit $y_*\in \{0,1,2,4,5,6,8,9\}$ (3 and 7 were omitted due to technical issues), for each scenario from Section~\ref{scenario_desciptions} we generate 50 repetitions of sequences $\{z_t\}$ of length 2,000.  The first 1,000 are $z_t$ where $\mathcal{M}$ is trained on all digits other than the omitted $y_*$.  In the last 1,000, $z_t$ are a mix of all digits including $y_*$, which is inserted in proportions $p_j$ corresponding to that scenario.  We use two CPM nonparametric tests, Student-T test (`\textbf{Student-T}'), which tests for changes in the mean confidence, and Cramer-von-Mises (`\textbf{CvM}'), which tests for arbitrary distribution change, each with $\alpha=0.05$.

In addition, for each, we try using the local density test (Section~\ref{sec:outlier_detection}) with $\alpha_*=0.05$, in which we do not return a CPM detection until at least one outlier is returned (that is, we have at least one $\{z_t\colon t\in \mathbf{\Theta\}}$ to present to the user.  This additional condition can only delay the detection relative, so the false alarm probability can only decrease but the missed alarm probability can increase if this prevents returning a detection.  These two tests are denoted `\textbf{Student-T\_outliers}' and `\textbf{CvM\_outliers}'.

Furthermore, we wish to demonstrate the fact that the CPM detectors are able to control the false alarm correctly (see Section~\ref{sec:sequential}), which methods that do not have sequential control do not.  Therefore, we compare our CPM results with two non-sequential methods:
\begin{itemize}
    \item `\textbf{naiveT\_pairwise}': calculate the p-value of the two-sample Student-T test between pairs of batches $\{z_1,\dots,z_{20}\}$ (the baseline) and $\{z_{20j-19}, z_{20j}\}$ for $j=2,\dots,100$; let the corresponding p-values be $\{\pi_2,\dots,\pi_{100}\}$ (not to be confused with the p-values of the local test in Section~\ref{sec:outlier_detection}).  Let $j_*=\text{min}(\{j\colon 2\leq j\leq 100\;\&\;\pi_j\leq \alpha\})$, that is the first batch $j$ that is declared significantly (defined by level $\alpha$) different from the initial one.  The best guess of changepoint is the previous batch, $\hat{B}=b(\hat{K})=j_*-1$.  If none are significant, $\hat{B}=\infty$. 
    \item `\textbf{naiveT\_splits}': This imitates the before/after splitting procedure of the CPM (see Algorithm~\ref{alg:cpm}).  At each batch $j\geq 2$, determine each before-after split at batch boundaries $\mathbf{Z}_{0;k,j}=\{z_1,\dots,z_{20k}\}$ and $\mathbf{Z}_{1;k,j}=\{z_{20k+1}\dots,z_{20j}\}$ for $1\leq k< j$.  Let the corresponding p-values for a two-sample Student-T test of $\mathbf{Z}_{0;k,j}$ vs $\mathbf{Z}_{1;k,j}$ be $\{\pi_{1,j},\dots,\pi_{j-1,j}\}$.  Let $\pi_j=\text{min}(\{\pi_{1,j},\dots,\pi_{j-1,j}\})$, the most significant split observed.  If $\pi_j\leq \alpha$, a change is declared and $\hat{B}=b(\hat{K})=j-1$, the previous batch.
\end{itemize}

Note that the non-sequential tests use the T-test without decision thresholds that account for the time elapsed, as opposed to the dynamic CPM thresholds $h_t$ (see Figure~\ref{fig:h_t}). 

In the following plots, we compare detection methods (CPM with/without outlier detection, and naive methods).  The results in each are aggregated across 50 repetitions each for each omitted digit class $y_*$.

\begin{figure}[ht]
    \centering
    \includegraphics[scale=0.4]{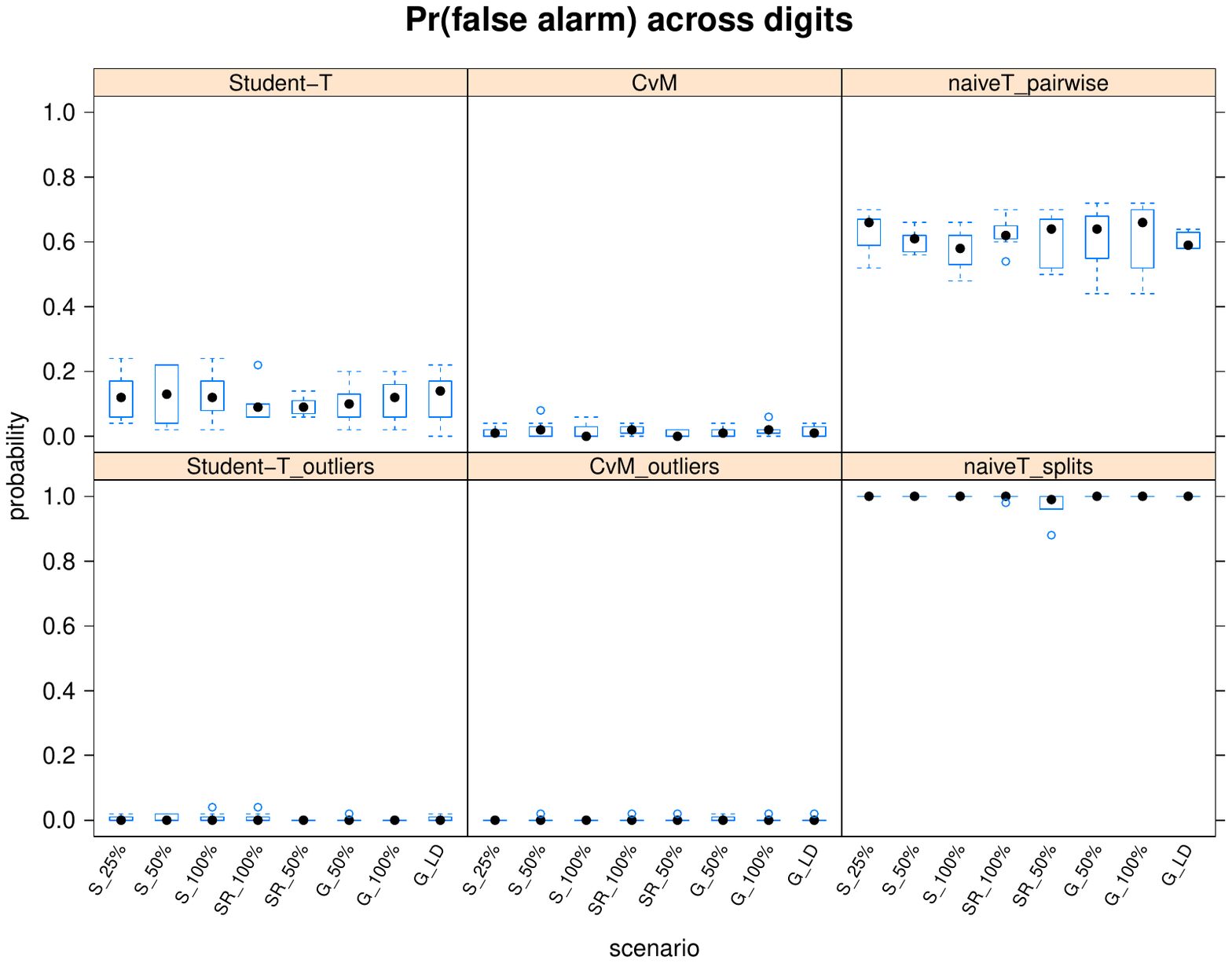}
    \caption{\label{fig:false_alarm} False alarm (Type-1 error) probabilities by scenario and detection method, aggregated across omitted digits and repetitions.}
\end{figure}

Figure~\ref{fig:false_alarm} shows the false alarm probabilities, that is, the probability of an early detection.  The Student-T CPMs (upper left corner) tend to have a false alarm rate of about 0.10 despite $\alpha=0.05$.  The CvM CPM (upper middle) has false alarm rates across all scenarios that are closer to the desired $\alpha$.  The non-sequential tests (right column) both have unacceptably-high false alarm rates, far above the 0.05 rate.  This demonstrates that non-sequential methods should not be blindly used if controlling the overall false alarm rate is desired.

\begin{figure}[ht]
    \centering
    \includegraphics[scale=0.4]{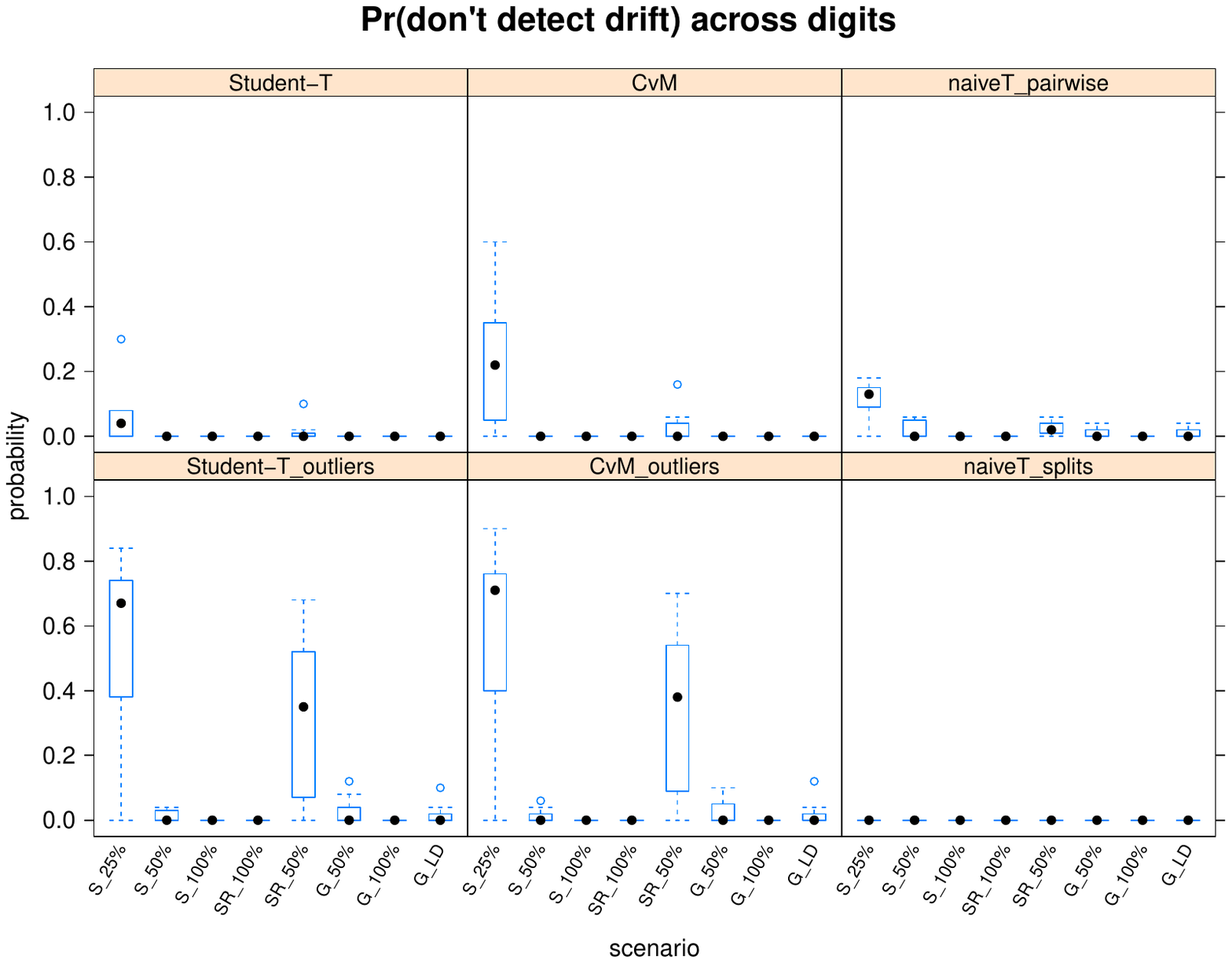}
    \caption{\label{fig:missed_detection}Missed alarm (Type-2 error) probabilities by scenario and detection method, aggregated across omitted digits and repetitions.}
\end{figure}

Figure~\ref{fig:missed_detection} shows the probability of not detecting change at all; this is the Type-2 error.  The true rate should be 0 since we do indeed simulate a distribution change, however it is possible that the detector would have detected if allowed more time.  By and large, in most scenarios and methods, the missed alarm rate is close to 0.  The scenarios \textbf{sudden\_quarter} (S\_25\%) and \textbf{sudden\_half\_return} (SR\_50\%) (see Figure~\ref{fig:drift_scenarios}) seem to be more difficult.  For the non-sequential tests, the low missed alarm rates are not much of a consolation since they have unacceptably-high false alarm rates.

\begin{figure}
    \centering
    \includegraphics[scale=0.4]{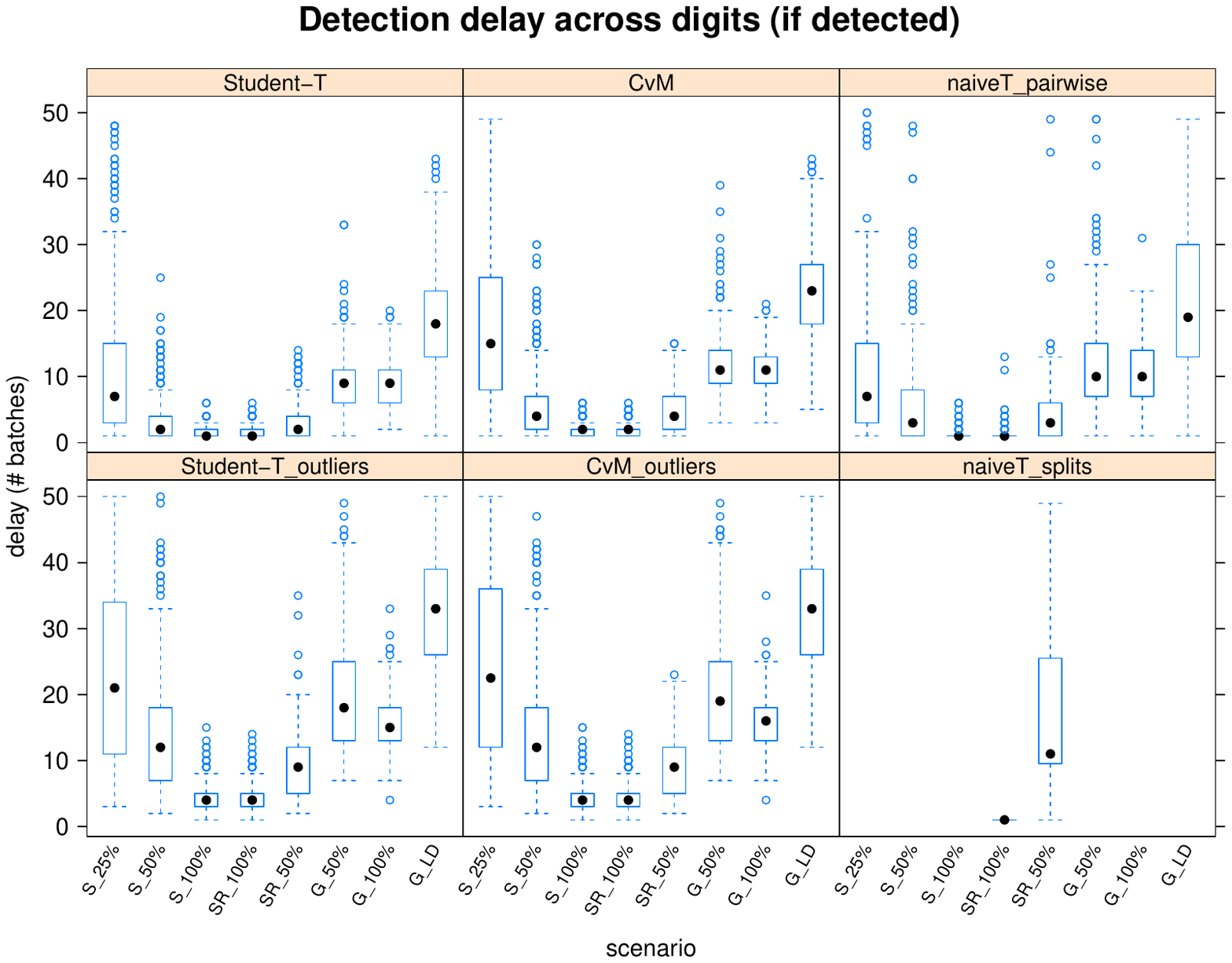}
    \caption{\label{fig:detection_delay}Batch detection delay ($b(d)-b(K)-1$), conditional on making a correct detection ($K<d<\infty$) by scenario and detection method, aggregated across omitted digits and repetitions.}
\end{figure}

Figure~\ref{fig:detection_delay} shows the detection delay by batch, conditional on making a detection.  The \textbf{naiveT\_splits} detector almost always makes a false alarm (almost no detections after $B=b(K)$ to consider for delays, see Figure~\ref{fig:false_alarm}), so most values are missing.  For the other cases, we see that the more gradual (G\_LD, G\_50\%, G\_100\%) scenarios are the hardest to detect quickly since the change is smallest.  S\_25\% also tends to have longer delays, even though it is a sudden change, since the contamination level is relatively low.  The sudden 100\% contamination scenarios (S\_100\% and SR\_100\%, in which the contamination disappears), have the shortest detection delays because the change is more abrupt.  Also, the delay for each CPM can only increase if the outliers requirement is added. 

\begin{figure}
    \centering
    \includegraphics[scale=0.4]{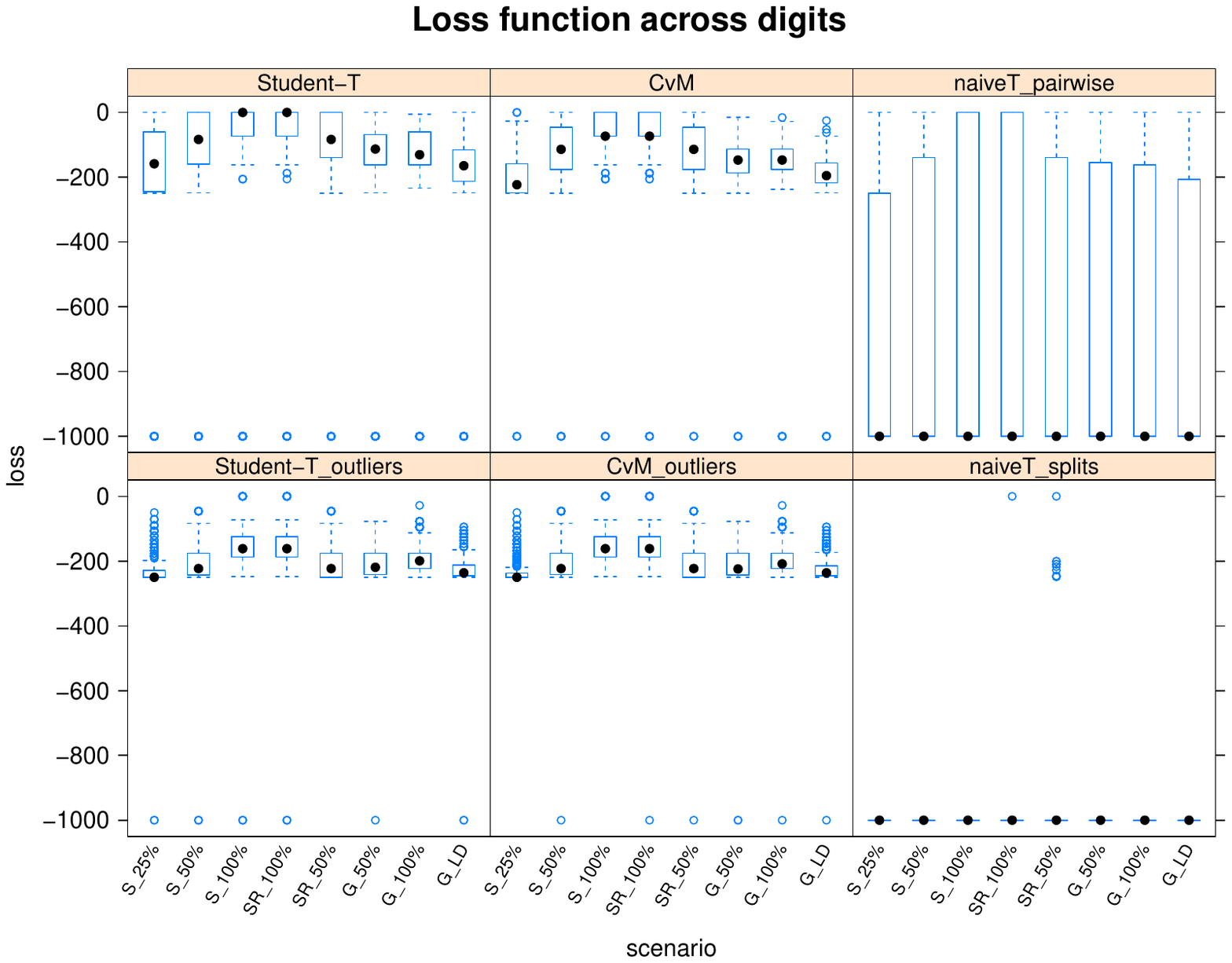}
    \caption{\label{fig:loss_across_digits}Loss $L(K,d,\{p_j\})$ by scenario and detection method, aggregated across omitted digits and repetitions.}
\end{figure}

In the previous figures, we have compared the methods across various scenarios. Scenarios with more abrupt change should have lower detection delay and also lower missed detection probability, but this relationship is likely not linear with the amount of contamination.  In Section~\ref{sec:scenarios} we introduced a loss function which takes into account the amounts of drift contamination $p_j$ and the delay (compounding each $p_j$ for the number of batches detection has been delayed on it) when assigning a score.  Boxplots of the loss functions are shown in Figure~\ref{fig:loss_across_digits}.  Since any false alarm yields a loss of $L_0=-1,000$, and a delayed or missed alarm yields a loss between $0$ and $L_1=-250$, the interesting part of the boxplot is between 0 and $-250$.  Note that in Figure~\ref{fig:drift_scenarios_loss}, the loss under all scenarios converges to nearly $L_1$, the loss for a missed detection, before the end of 100 batches, so the boxplots in this area will not be skewed by missed detections.

For instance, in Figure~\ref{fig:detection_delay} for  the CvM CPM (middle upper row), consider the rightmost two scenarios, G\_100\% and G\_LD; the inter-quartile ranges (rectangles) of the two boxplots do not even overlap, indeed the whole distribution of delay for G\_100\% is lower than the mean of that of G\_LD.  However, the delayed detection of G\_LD is justified because its contamination is much more gradual.  In Figure~\ref{fig:loss_across_digits}, the loss for G\_LD is more negative (worse) than G\_100\%, but not worse to the same degree as is its comparative detection delay, since the lower drift contamination in G\_LD is penalized less than is the sudden full contamination in G\_100\%.  The loss function at least allows us to compare performance under different scenarios on a more level ground.

In Section~\ref{sec:outlier_detection} we stated an additional goal of returning particularly anomalous data examples to the user if model performance change is detected, to allow them to better diagnose the cause of the change by manual inspection or simple analysis.  We used a local density-difference test to determine indices $t\in \mathbf{\Theta}$ corresponding to observed confidences $\{z_t\}$ in the determined post-change sample $\mathbf{Z}_1$ (whose density is estimated by $\hat{f_1}$), which appear anomalous relative to the pre-change sample $\mathbf{Z}_0$ (whose density is estimated by $\hat{f_0}$.  Here, we select up to 10 of the most anomalous cases ($1\leq |\mathbf{\Theta}|\leq 10$).  The goal is that $\theta :=\displaystyle{\frac{\sum_{t\in \mathbf{\Theta}}I(y_t=y_*)}{|\mathbf{\Theta}|}}$\;---that is, the fraction of returned outliers $\{\mathbf{x}_t\}$ that are the drift class (omitted digit)---will be relatively high, meaning we are relatively confident our outliers are of the drift class.   

In Figure~\ref{fig:detection_delay}, let us compare the left column panels (CPM without requiring an outlier to be returned) with the middle column panels (delaying change detection until at least one outlier, subject to threshold $\alpha_*$).  Requiring outliers to be found---whether or not they are of the drift class---delays detection somewhat; the degree depends on the scenario. 

\begin{figure}
    \centering
    \includegraphics[scale=0.4]{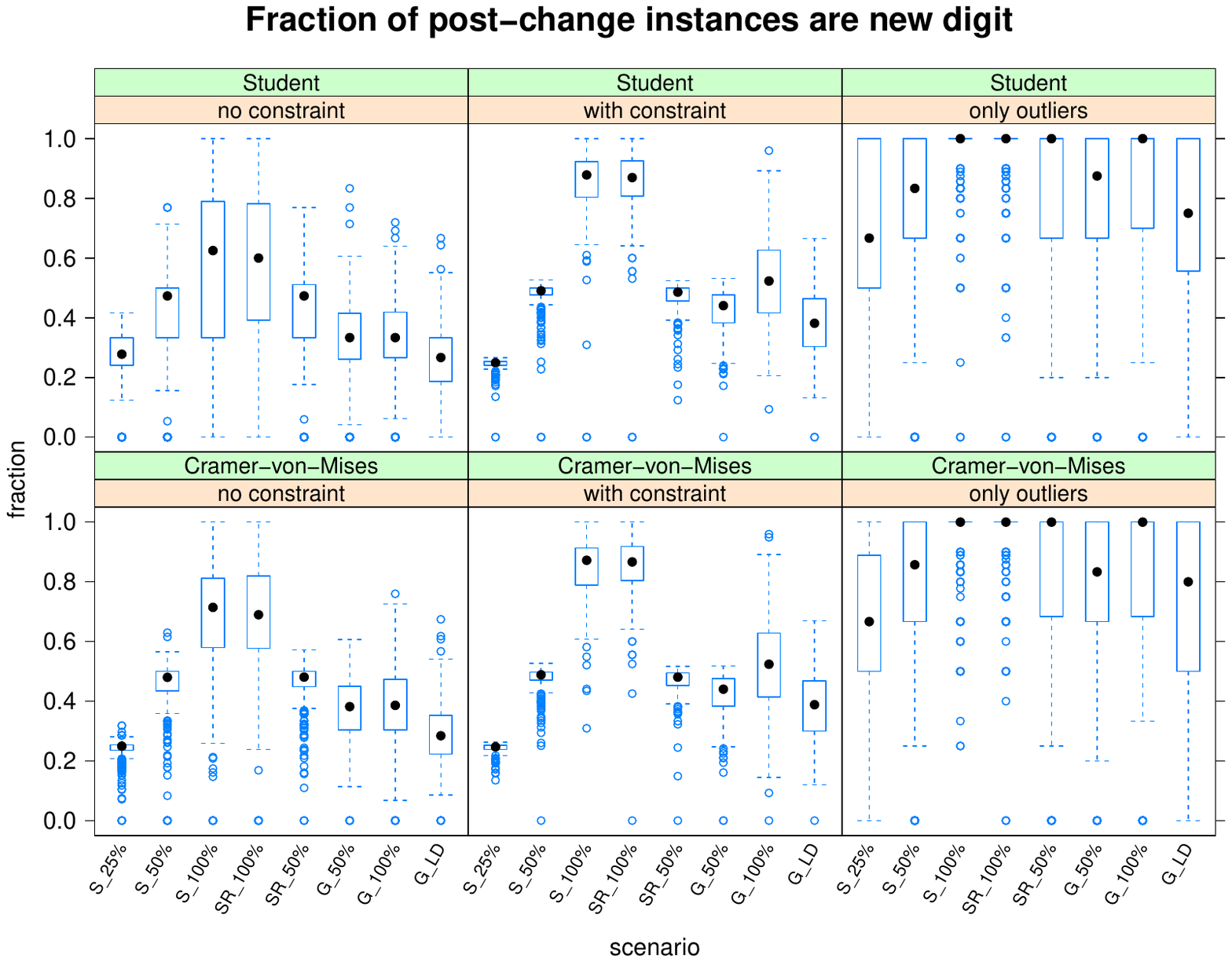}
    \caption{\label{fig:outlier_drift_frac}Probability of identified post-change sample that is the drift fraction (denoted $\theta$), for the two CPM methods. Left column is CPM without testing for outliers; middle column is CPM with testing for outliers, but before most-significant outliers are selected; right column is only the most significant outliers.}
\end{figure}

In Figure~\ref{fig:outlier_drift_frac}, we demonstrate the success of our method by examining the empirical distribution of the fraction $\theta$.  The left column (``no constraint") shows, for Student-T and CvM CPM (`\textbf{Student-T}' and `\textbf{CvM}' of the previous figures), the fraction of the post-change sample $\mathbf{Z}_1$ that are examples of the drift class (omitted digit).  That is, we do not apply an outlier test, but simply let $\mathbf{\Theta}=\{\hat{K}+1,\dots,d\}$, that is, all indices after the estimated change point $\hat{K}$, when calculating the fraction $\theta$.  The middle column (``with constraint") has $\theta$ similarly defined, where we show the results where $\mathbf{\Theta}=\{\hat{K}+1,\dots,d\}$ when we wait for the test to identify at least one outlier, but before we select only outliers in $\mathbf{\Theta}$.  In order for an outlier to be identified after the CPM detects change, the overall drift fraction $\frac{\sum_{j=\hat{K}+1}^dp_j}{d-\hat{K}}$ typically needs to be higher than without the outlier requirement, since a higher $\theta$ will increase the number of potential outliers.  Thus, the average drift contamination in the middle column is higher than in the first, since in our scenarios $p_j$ tends to increase with the batch $j$.  

Let $\mathbf{\Theta}=\{\hat{K}+1,\dots,d\}$ be the indices after the estimated changepoint when we find at least one outlier, but before selecting the outliers.  Let $\mathbf{\Theta}^{'}$ be the (up to 10) indices sampled from $\mathbf{\Theta}$ that are the most anomalous (as defined by our local density test).  To be successful, we need $\theta^{'}$ defined on the set $\mathbf{\Theta}^{'}$ to be larger than the average fraction $\theta$ calculated on the set $\mathbf{\Theta}$.  That is, we need to show our outlier detection method actually finds a significantly higher fraction of drift class (indirectly) than would taking a simple random fraction of post-change observations indexed $\mathbf{\Theta}$, which has average drift contamination $\theta$.  The right column (``only outliers") of Figure~\ref{fig:outlier_drift_frac} shows these fractions $\theta^{'}$; the middle column are the corresponding $\theta$ without the test.  As we can see, the boxplots in the right column are significantly higher than the corresponding ones in the middle column.  Indeed, many of them return an average of 60\% or above drift class, meaning that the returned set of outliers should contain at least several drift instances that should be easy to identify.  Furthermore, these fractions $\theta^{'}$ are high \textit{across} scenarios, thought they are highest on average in the more abrupt scenarios S\_100\% and SR\_100\%. 

\section{Conclusion
\label{sec:conclusion}}
In this work, we discussed a method to statistically detect changes in the performance of a ML model $\mathcal{M}$ by observing changes in the distribution of its prediction confidence $z_t$ over time.  It is often the case the the performance (as measured by $z_t$) can change, particularly for the worse, when it encounters data observations $\mathbf{x}_t$ which may be unusual in some way from $\mathcal{M}$'s training data.  Our method uses the CPM (\cite{RA12}) a nonparametric method that is both more statistically correct than other methods, and allows us to indirectly infer changes in model performance regardless of the data domain.

We tested our method by omitting a different MNIST digit class from the training set each time, and seeing if we could detect the omitted class in the test (deployment) phase.  In our experiments, we simply relied on detecting changes in the prediction confidence observation without `telling' our detector what type of data change it expected to find.  Our results (see Section~\ref{sec:results}) show that we are able to reliably detect these changes while controlling for false alarms, and additionally present potentially anomalous data instances (see Figure~\ref{fig:outlier_drift_frac})---again, without defining the type of anomaly directly---that have a high likelihood of belonging to the drift class (here, the omitted digit).  

In theory, our approach should work on performance metrics other than confidence, as long as these metrics are sensitive to changes in the data (e.g., new class).  We note that confidence does not require knowledge of the ground truth $y_t$, whereas other metrics like F-score and mean squared error (MSE) do.  On one hand, if the metric is sensitive to data changes (`drift'), we can use changes in the performance to identify drift.  But, if the performance metric does not change, perhaps this is good for the model $\mathcal{M}$ because it means $\mathcal{M}$ is robust to data drift according to this metric.  In the case we have experimented with, where the drift takes the form of an unseen class, a traditional classifier cannot classify it correctly since it only knows the set of labels (digits) it was trained on.  In this classification task, it would not help if the prediction confidence was robust to the new class, since $\mathcal{M}$ could never output the correct label; in fact, this would be a bad property.  In a numeric prediction task, such as linear regression, however, a model could still have strong performance (e.g., low MSE) despite encountering drift in the form of values outside the expected range, if the modeled relationship still holds. A sudden change in the MSE could indicate drift in the form of a change in the model relationship.  Thus, a detector which used MSE could detect the second kind of drift, which is more problematic, but not the first.  

%Note:BibTeX also works

% \begin{references}
% \itemsep=0pt
% {\footnotesize
% \bibliographystyle{plain}

% }
% \end{references}
\itemsep=0pt
{\footnotesize
\printbibliography

@article{LMD11, 
author = {Patrick Lindstrom and Brian Mac Namee and Sara Jane Delaney},
title = {Drift Detection Using Uncertainty Distribution Divergence}, 
journal={Evolving Systems},
year = 2011,
vol= 4,
pages= {13–25},
}

@article{L99,
author = {Carsten Lanquillon},
year = {1999},
month = {06},
pages = {},
title = {Information Filtering in Changing Domains},
journal = {Proceedings of the International Joint Conference on Artificial Intelligence}
}

@article{D13,
author = {Tarn Duong},
year = {2013},
title = {Local Significant Differences from Non-Parametric Two-Sample Tests},
journal = {Nonparametric Statistics}
}

@misc{DWCG18,
author = {Tarn Duong and Matt Wand and Jose Chacon and Artur Gramacki},
year = {2018},
title = {ks: Kernel Smoothing},
journal = {R package},
url={https://cran.r-project.org/web/packages/ks/index.html}
}

@article{RA12,
author = {Gordon J. Ross and Niall M. Adams},
year = {2012},
title = {Nonparametric Control Charts for Detecting Arbitrary Distribution Changes},
journal = {Journal of Quality Technology}
}

@article{KR98,
author = {Ralf Klinkenberg and Ingrid Renz},
year = {1998},
title = {Adaptive Information Filtering: Learning in the Presence of Concept Drifts},
journal = {AAAI Technical Report}
}

@Article{MRACH12,
    title = {A Unifying View on Dataset Shift in Classification},
    author = {Jose G. Moreno-Torres and  Troy Raede and Roc\'{i}o Alaiz-Rodr\'{i}guez and Nitesh V. Chawla and Francisco Herrera},
    journal = {Pattern Recognition},
    year = {2012},
    volume = {45},
    pages = {521--530}
  }

@Article{FDA20,
    title = {Sequential Drift Detection in Deep Learning Classifiers},
    author = {Samuel Ackerman and Eitan Farchi and Parijat Dube},
    journal = {arXiv},
    url={https://arxiv.org/pdf/2007.16109.pdf},
    year = {2020},
  }

@Article{AFRZZ19,
    title = {Automatically detecting data drift in machine learning classifiers},
    author = {Samuel Ackerman and Eitan Farchi and Orna Raz and Marcel Zalmanovici and Aviad Zlotnick},
    journal = {Association for the Advancement of Artificial Intelligence},
    year = {2019},
  }

@misc{LCB99,
    title = {The MNIST Handwritten Digit Database},
    author = {Yann LeCun and Corinna Cortez and Christopher C.J. Burges},
    url={http://yann.lecun.com/exdb/mnist/},
    year = {1999},
  }

@Article{RATH12,
    title = {Exponentially Weighted Moving Average Charts for Detecting Concept Drift},
    author = {Gordon J.Ross and Niall M. Adams and Dimitris K. Tasoulis and David J. Hand},
    journal = {Pattern Recognition Letters},
    year = {2012},
    vol={33},
    pages={191--198},
  }

@misc{R15,
author = {Gordon J. Ross},
year = {2015},
title = {cpm: Sequential and Batch Change Detection Using Parametric and
Nonparametric Methods},
journal = {R package},
url={https://cran.r-project.org/web/packages/cpm/index.html}
}

@Article{SK17,
    title = {On the Reliable Detection of Concept Drift from Streaming Unlabeled Data},
    author = {Tegjyot Singh Sethi and Mehmed Kantardzic},
    journal = {Expert Systems with Applications},
    year = {2017},
    vol={82},
    pages={77--99},
    url = {https://www.sciencedirect.com/science/article/abs/pii/S0957417417302439},
  }

@Article{M56,
    title = {The Magical Number Seven, Plus or Minus Two: Some Limits on Our Capacity for Processing Information},
    author = {George A. Miller},
    journal = {Psychological Review},
    year = {1956},
  }
}
%\printbibliography

\end{document}